\documentclass[aps,prl,reprint,superscriptaddress]{revtex4-1}

\usepackage{longtable}
\usepackage{morefloats}
\usepackage{color}
\usepackage{graphicx,amsfonts,amsbsy}
\usepackage{amsmath,amsfonts,amsthm,amssymb}
\usepackage{appendix}
\usepackage{makeidx}
\usepackage{url}
\usepackage{verbatim}
\usepackage[bookmarksnumbered,pdfpagelabels=true,plainpages=false,colorlinks=true,linkcolor=blue,citecolor=blue,urlcolor=blue]{hyperref}
\usepackage[rightcaption]{sidecap}
\usepackage{array}
\usepackage{booktabs}
\usepackage{multirow}
\usepackage{tabularx}
\usepackage{cancel,soul,ulem}

\usepackage{mathrsfs}

\newcommand{\bs}[1]{\boldsymbol{#1}}

\begin{document}
\title{Time-dependent strain-tuning topological magnon phase transition}

\author{Nicolas Vidal-Silva}
\affiliation{Departamento de Ciencias F\'isicas, Universidad de La Frontera, Casilla 54-D, Temuco, Chile}

\author{Roberto E. Troncoso}
\affiliation{Center for Quantum Spintronics, Department of Physics, Norwegian University of Science and Technology, NO-7491 Trondheim, Norway}

\begin{abstract}
Collinear magnets in honeycomb lattices under the action of time-dependent strains are investigated. Given the limits of high-frequency periodically varying deformations, we derive an effective Floquet theory for spin system that results in the emergence of a spin chirality. We find that the coupling between magnons and spin chirality depends on the details of the strain such as the spatial dependence and applied direction. Magnonic fluctuations about the ferromagnetic state are determined, and it is found that spatially homogeneous strains drive the magnon system into topologically protected phases. In particular, we show that certain uniform strain fields play the role of an out-of-plane next-neighbor Dzyaloshinskii-Moriya interaction. Furthermore, we explore the application of nonuniform strains, which lead to a confinement of magnon states that for uniaxial strains, propagates along the direction that preserves translational symmetry. Our work demonstrates a direct and novel way in which to manipulate the magnon spectrum based on time-dependent strain engineering that is relevant for exploring topological transitions in quantum magnonics.
%
\end{abstract}

\maketitle
Magnons---the elementary bosonic excitations of the magnetic order---are a key ingredient for
future spin-based, low-power-consumption and ultralow-noise technologies \cite{chumak2015magnon}. The control and transport of magnon spin currents over large distances, e.g., in diffusive regimes or in superfluid systems \cite{cornelissen2015long,QaiumzadehPRL2017}, constitute one of the main challenges of magnonics \cite{barman20212021}.

The interest in spin fluctuations in magnetic insulators was invigorated by the discovery of exotic topological phenomena \cite{mcclarty2021topological,wang2018topological,wang2021topological,malki2019topological,zhu2021topological,ghader2020magnon,wang2020bosonic}. These effects have their roots in the geometric properties of the space of magnonic eigenstates \cite{wang2017topological}, which result in the emergence of exceptional phenomena such as the magnon Hall effect \cite{onose2010observation} and robust edge-modes \cite{mook2014magnon}. The realization of topological spin excitations has been shown in various two-dimensional honeycomb van der Waals magnets, either intrinsic \cite{zhu2021topological,aguilera2020topological,li2019intrinsic,hidalgo2020magnon,pershoguba2018dirac,qin2019emergence,ghader2022theoretical} or induced by light \cite{owerre2017floquet,vinas2020light}. This phenomenology, either in collinear or textured magnetic materials, is related to a (scalar) spin chirality,  $\chi_{ijk}={\bs S}_j\cdot \left( \bs{S}_i \times \bs{S}_{k}\right)$, which appears as an emergent magnetic field and is responsible for the topological transport of magnons \cite{lu2019topological,lee2015thermal,katsura2010theory,han2017spin,kanazawa2011large,taguchi2001spin,onoda2002topological}.

{Mechanical strains are a widely recognized technique to create artificial gauge fields in solid systems \cite{vozmediano2010gauge,amorim2016novel}.
This method enables the engineering of states of matter such as Landau levels in strained graphene \cite{lantagne2020dispersive,uchoa2013superconducting,li2020valley,hsu2020nanoscale} or topological phase transitions \cite{liu2014tuning,zeljkovic2015strain,zhang2018topological,liu2016strain,munoz2022inducing}. Similarly, elastic gauge fields have been predicted in magnetic insulators through spatial modulations of the exchange interaction. The application of suitable nonuniform and stationary strain patterns results in novel magnonic states. A few strained magnetic systems have been studied in this context, including topological phases in ferromagnetic Kagome lattices \cite{OwerreJPCM2018}, magnon (pseudo) Landau levels in ferromagnets \cite{FerreirosPRB2018,LiuPRB2021} and in antiferromagnets \cite{SunPRB2021,SunPRR2021}, which have also shown exotic emergent supersymmetry properties under triaxial strains \cite{NaygaPRL2019}.

In this Letter, we propose a mechanism of time-dependent strain engineering to induce an effective spin chirality in the spin system. We show that high-frequency periodically varying uniform strains modify the spectrum of magnonic excitation into topologically gapped magnon states. Moreover, when the spatial component of the strain is nonuniform, it breaks translational symmetry and gives rise to a confinement of magnon states.\begin{figure}[ht]
		\includegraphics[width=85mm]{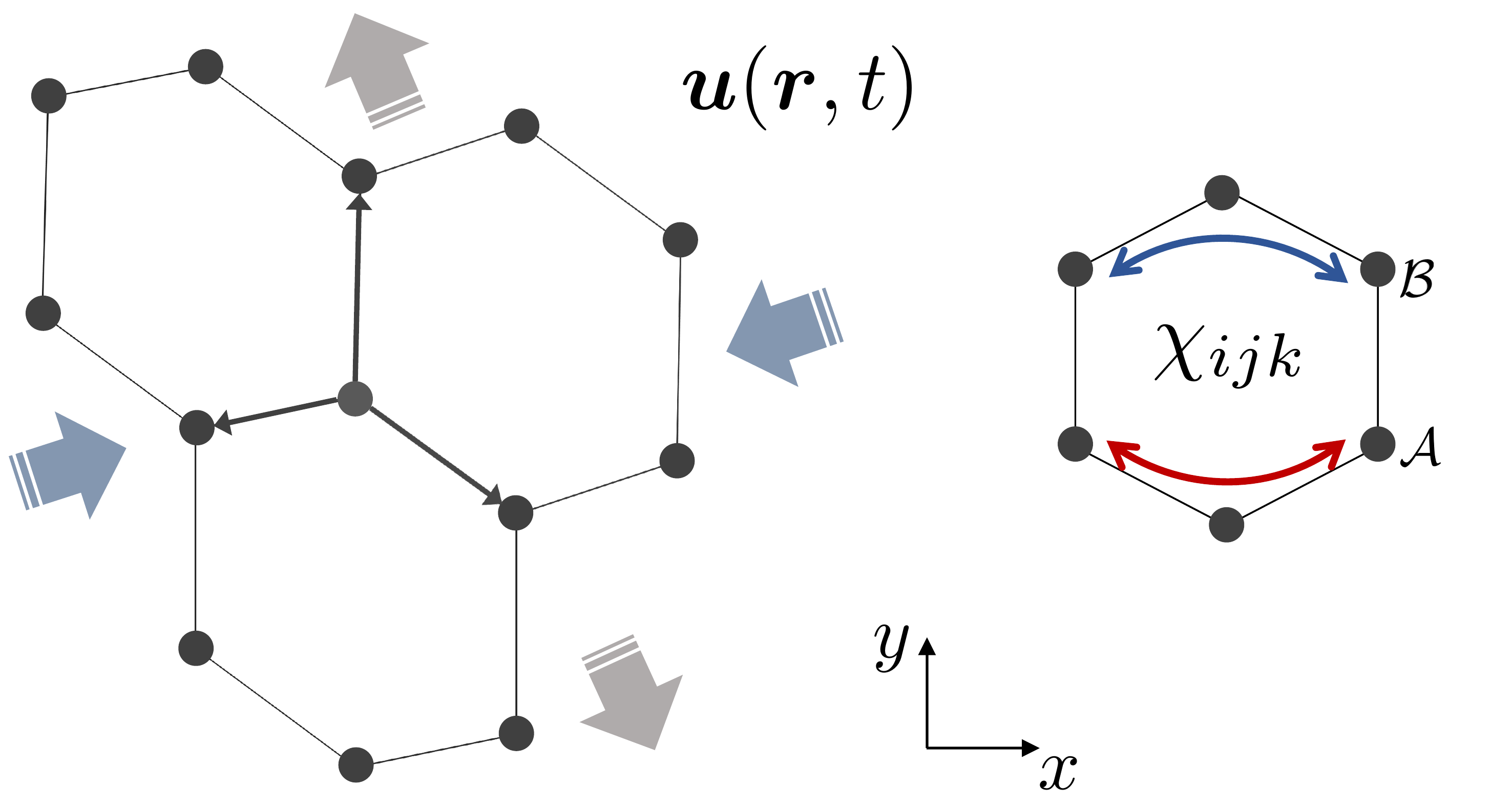}
		\caption{Schematic illustration of deformed honeycomb lattices. Time-dependent strains are described by in-plane deformation field, ${\bs u}({\bs r},t)$. Effective spin chirality $\chi_{ijk}$, relating next-nearest neighbor spins, emerges from high-frequency strain fields.}
\label{fig: hex-lattice-deformed}
\end{figure}
The underlying mechanism is substantially different from those systems under applied static strains, where the notion of an elastic gauge field emerges for smooth deformations and near the Dirac points \cite{vozmediano2010gauge,amorim2016novel}. Instead, periodically varying strains, treated in perturbation theory, manifest as a massive term in continuum magnon theory, leading to several consequences on magnon properties. This result, obtained in the high-frequency limit, is a key difference with respect to emergent pseudo electric fields that lead to topological responses \cite{sela2020quantum,bhat2018time,vaezi2013topological} and are induced by slowly varying mechanical deformations.}

{\it Model}.-- We consider a system of localized spins placed on a two-dimensional honeycomb lattice. The nearest-neighbor spin Hamiltonian that allows for applied mechanical strains is
\begin{align}\label{eq:spinHamiltonian}
{H}=-\sum_{\langle ij\rangle}{\frac{J_{ij}}{2}}{\bf S}_i\cdot{\bf S}_j-\frac{\cal {K}}{2}\sum_i({S}^z_{i})^2-B\sum_i{S}^{z}_{i},
\end{align}
where $J_{ij}\equiv J({\bs r}_i-{\bs r}_{j})>0$ is the ferromagnetic exchange coupling between spins located at positions ${\bs r}_i$ and ${\bs r}_j$ in the lattice. The strength of the easy-axis magnetic anisotropy is ${\cal K}$, and $B$ is the external magnetic field. The elastic deformations conducted by mechanical strains are represented by the field ${\bs u}({\bs r})$ describing the displacement of atomic sites from equilibrium positions. The strain is coupled to the system of spins through the magnetoelastic interaction, which adopts a simple form via spatial modulations of the exchange coupling as $J_{ij}\approx J-J'{\bs \delta}_{\eta}\cdot({\bs \delta}_{\eta}\cdot\nabla){\bs u}({\bs r}_i)$ for small in-plane displacements \cite{Neek-Amal-PRB2013},being $J$ the exchange coupling for the unstrained system and the nearest neighbor vectors ${\bs\delta}_{\eta}$ are defined in Fig. \ref{fig: hex-lattice-deformed}. In the following, we assume that a collinear and out-of-plane ferromagnetic order is preserved in the strained system.

{\it Floquet theory and effective Hamiltonian}.-- We now consider the spin system driven by time-dependent strains. The Hamiltonian for the strained system is written as $H(t)=H+V(t)$, with $V(t)=\sum_{\langle ij\rangle}{\cal J}_{ij}(t)\,{\bf S}_i\cdot{\bf S}_{j}$ being the time-periodic potential. In terms of the strain tensor $\epsilon_{\alpha\beta}=(\partial_{\alpha}u_{\beta}+\partial_{\beta}u_{\alpha})/2$, the magnetoelastic coupling is written as ${\cal J}_{ij=i+{\bs \delta}_{\eta}}=J'{\delta}^{\alpha}_{\eta}{\delta}^{\beta}_{\eta}\epsilon_{\alpha\beta}({\bs r}_i,t)$, where $\alpha,\beta \in \{x,y\}$ and summation over repeated indexes is implicit. In addition, we represent the time-dependent field of deformations as ${\bs u}({\bs r}_i,t)={\bs u}^a({\bs r}_i)\sin \omega t+{\bs u}^b({\bs r}_i) \cos{\omega t}$, with ${\bs u}^a$ and ${\bs u}^b$ nonparallel vectors. In the limit of high-frequency strains, an effective model results in the time-averaged periodic Hamiltonian. Within the framework of Floquet-Magnus expansion \cite{MikamiPRB2016,MohanPRB2016}, the effective theory is obtained perturbatively, including corrections of ${\cal O}\left(\omega^{-n}\right)$, and is quadratic on the magnetoelastic coupling. This results in the effective Hamiltonian $H_{\text{eff}}= H+\sum_n{\left[H_n,H_{-n}\right]}/{n\omega}$, where $H_n=1/{\cal T}\int^{\cal T}_0 dt H(t)e^{in\omega t}$ is the Fourier component of the time-dependent Hamiltonian and ${\cal T}=2\pi/\omega$ is the period of the deformation field. Therefore, we obtain
\begin{align}\label{eq:effecHam}
H_{\text{eff}}=H+\frac{2i}{\omega}\sum_{\langle ij\rangle}\sum_{\langle jk\rangle}\Delta_{ijk}\,
{\bs S}_j\cdot \left( \bs{S}_i \times \bs{S}_{k}\right),
\end{align}
with ${\bs S}_j=S_j^z\bs{\hat{z}}$. The second term on the right-hand side of Eq. (\ref{eq:effecHam}) corresponds to the strain-induced spin-spin interaction that constitutes the central result of this Letter.
The coupling tensor $\Delta_{ijk}={\cal J}^{+}_{ij}{\cal J}^{-}_{jk}-{\cal J}^{+}_{jk}{\cal J}^{-}_{ij}$, where ${\cal J}^{\pm}_{ij}=({\cal J}^{b}_{ij}\pm i{\cal J}^{a}_{ij})/2$, is nonlocal and quantifies the effective interaction of a triad of next-neighboring spins (see Fig. \ref{fig: hex-lattice-deformed}  and the Supplemental Material for major details). In general, the time-dependent strain induces spin chirality ${\bs S}_i\cdot\left({\bs S}_j\times{\bs S}_k\right)$, which modifies the static Hamiltonian. As a result, the energy landscape for the magnetic order might lead to the stabilization of magnetic textures, e.g., skyrmions \cite{nagaosa2013topological}. However, to understand the effective Hamiltonian $H_{\text{eff}}$, we will focus on the analysis of linear spin fluctuations. It is important to remark that the strain-induced Floquet correction is independent of the actual magnetic order and therefore is also valid for antiferromagnetic models.

{\it Strained magnon Hamiltonian}.-- We now study the low-energy spin fluctuations of the strain-induced effective Hamiltonian within linear spin-wave theory. It is convenient to introduce bosonic operators through the Holstein-Primakoff (HP) formalism \cite{holstein1940field}. The ordered magnetic ground state is assumed to be out-of-plane; therefore, the quantization axis is along the $z$-axis. Thus, at lattice site $i$, the spin operators and HP bosons are related by $s^-_i=a^{\dagger}_i\sqrt{2s-a^\dag_i a_i}$, $s^+_i=\sqrt{2s-a^\dag_i a_i}a_i$ and $s^z_i = s - a^\dag_i a_i$. We expand the spin Hamiltonian Eq. (\ref{eq:effecHam}) in terms of HP bosons and disregard many-body magnon interactions. Introducing the field operator $\Psi_{\bs k}=(a_{\bs k},b_{\bs k})^T$, with $a$ and $b$ the bosonic operators defined on sublattices ${\cal A}$ and ${\cal B}$, respectively, we find for the momentum space effective magnon Hamiltonian,
\begin{align}\label{eq: magnonHamiltonian}
H_m=\sum_{{\bs k},{\bs k}'}\Psi^{\dagger}_{\bs k}\left(\Omega\tau_0\,\delta_{{\bs k}{\bs k}'}+{\bs h}_{\bs k\bs k'}\cdot{\bs{\tau}}\right)\Psi_{\bs k'},
\end{align}
where $\bs{\tau}$ is a pseudovector of the Pauli matrices, $\tau_0$ is the identity matrix, and $\Omega=sJz+sK+B$, where $z$ is the coordination number.
\begin{align}\label{eq: h-field}
{\bs h}_{\bs k\bs k'}=
\left(\begin{array}{c}
-sJ\sum_{\eta}\cos\left[{\bs k}\cdot{{\bs\delta}_{\eta}}\right]\delta_{{\bs k}{\bs k}'}\\
sJ\sum_{\eta}\sin\left[{\bs k}\cdot{{\bs \delta}_{\eta}}\right]\delta_{{\bs k}{\bs k}'}\\
4s^2(J')^2{\Delta}_{\bs k\bs k'}/{\omega}
\end{array}\right).
\end{align}
The field $\Delta_{{\bs k}{\bs k}'}
=2i\sum_{\eta}\Delta_{{\bs k}{\bs k}'}^{\eta}\sin\left(\bs{k}'\cdot\bs{\delta}^{nn}_{\eta}\right)/N$, with $\Delta_{{\bs k}{\bs k}'}^{\eta}=\sum_{i}\Delta_{i,i-\bs{\delta}_{\eta},i-\bs{\delta}^{nn}_{\eta}} e^{-i({\bs k}-{\bs k}')\cdot{\bs r}_i}$,$N$ the number of lattice sites, and the next nearest-neighbor vectors ${\bs \delta}^{nn}_{\eta}$ (see Fig. \ref{fig: hex-lattice-deformed} and Supplemental Material) features the nonlocality of magnon coupling in momentum space.

There are two remarkable characteristics of the magnon Hamiltonian (\ref{eq: magnonHamiltonian}). First, high-frequency periodically driven strains emerge as a massive term (${h}^z_{\bs k\bs k'}$) in the theory for noninteracting magnonic fluctuations. This is quite different from systems under the action of static lattice distortions. There, nonuniform strains manifest as elastic gauge fields close to the Dirac points, modifying the in-plane components of ${\bs h}_{\bs k\bs k'}$ and giving rise to pseudo-Landau levels in the energy spectrum \cite{OwerreJPCM2018,FerreirosPRB2018,LiuPRB2021,SunPRB2021,SunPRR2021,NaygaPRL2019}. Second, due to strain-induced coupling $\Delta_{\bs k\bs k'}$, the effective theory for magnons becomes nonlocal in momentum space. The dependence on momenta ${\bs k}$ and ${\bs k}'$ is established by the spatial dependence of the strain tensor. In particular, for homogeneous deformations of the lattice, ${\bs u}({\bs r})={\bs u}_0$, the field $\Delta_{ijk}$ becomes null, similar to the strain tensor. Thus, any nontrivial effect on the system of magnons is expected for spatially dependent lattice deformations.

To study the spectra of magnonic excitations, we consider lattice deformations induced by homogeneous and nonuniform mechanical strains. While the former derives from linearly varying deformation fields ${\bs u}_l({\bs r})$, the latter originates from any field of displacements ${\bs u}_{nl}({\bs r})$ that depends on position in a nonlinear fashion. In the first case, elastic deformations are parameterized by two time-dependent fields with amplitudes of oscillation $\bs{u}^a_l$ and $\bs{u}^b_l$. We find that the strain-induced magnon coupling becomes local in momentum space, ${\Delta}_{\bs k\bs k'}={\Delta}_{\bs k}\delta_{\bs k\bs k'}$, with ${\Delta}_{\bs k}=-{ a^4} \sum_{\eta}\Delta^{(\eta)}\sin\left(\bs{k}\cdot\bs{\delta}^{nn}_{\eta}\right)$ and $\Delta^{\eta}$ quantifying the effective next-nearest neighbor interaction. In particular, for the specific lattice deformations, $\bs{u}^a_l=(a_2y,b_1x)$ and $\bs{u}^b_l=c_1(x,-3y/2)$, with $a_1$, $b_1$ and $c_1$ being constants (see Supplementary material), the coupling is identical for each bond and $\Delta^{(\eta)}\equiv\Delta=3\sqrt{3}c_1(a_2+b_1)/2$. This is equivalent to a Dzyaloshinskii-Moriya interaction with strength $D_{\text{eff}}=-{4s^2a^4(J')^2\Delta}/{\omega}$ ($a$ the lattice parameter) that results from inversion symmetry breaking and is responsible for opening a topological gap at the Dirac points \cite{mcclarty2021topological,wang2018topological,wang2021topological,malki2019topological}.
\begin{figure}[ht]
\includegraphics[width=85mm]{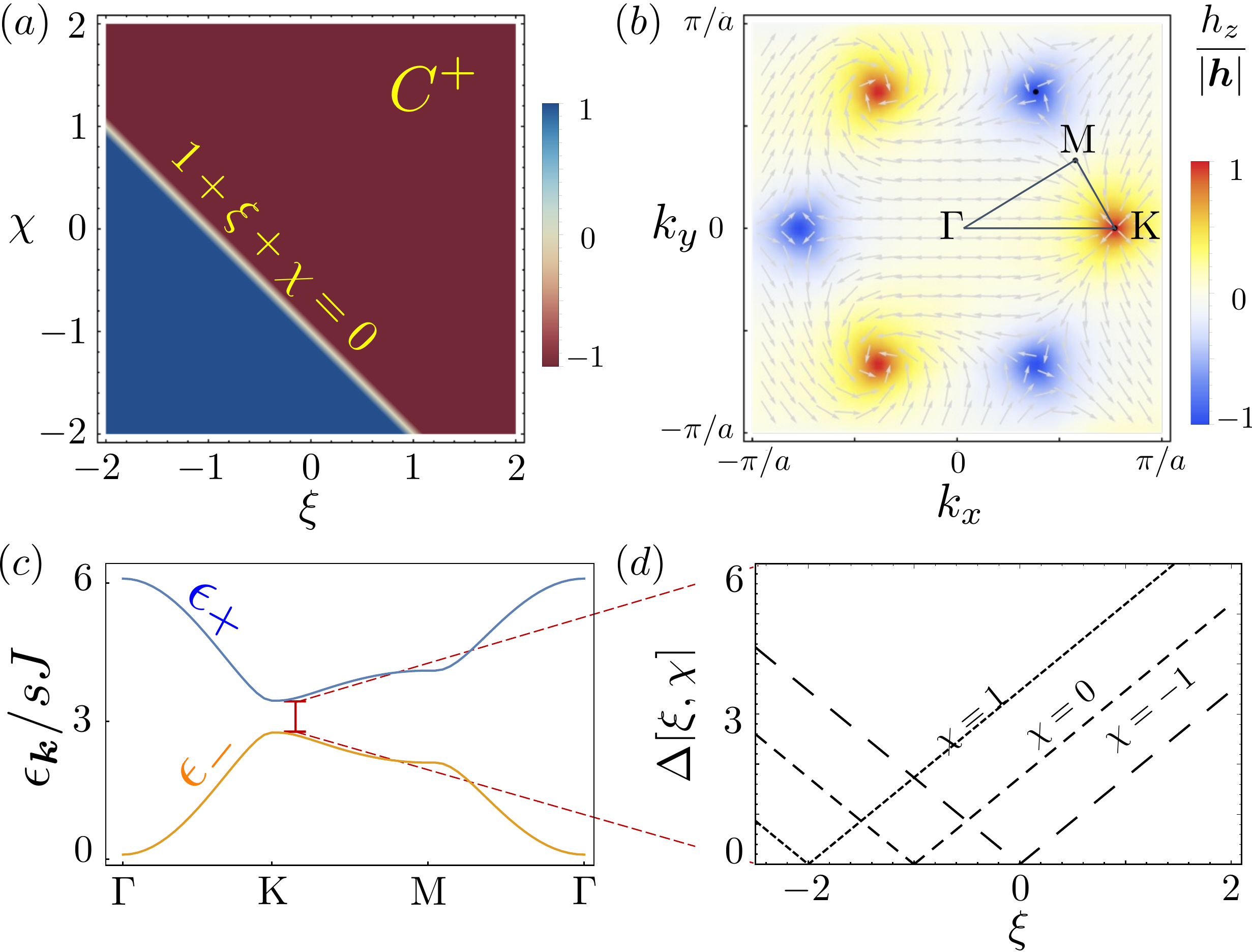}
\caption{(a) Topological phase diagram exhibited by strained magnon system in terms of parameters $\xi$ and $\chi$. Transition (topological) is set by sign change of Chern number, which is determined by condition $1+\chi+\xi=0$. Chern numbers for upper and lower magnonic modes are related by $C^+=-C^-$.
(b) Effective field ${h}_z/|{\bs h}|$ is shown in momentum space for specific values $\chi=2$ and $\xi= 1$. Vector field is given by the $x$ and $y$ components of normalized field ${\bs h}/|{\bs h}|$. For same values of $\chi$ and $\xi$, dispersion relation is displayed in (c). Magnon energy bands $\epsilon_{\pm,\bs k}$ are gapped at Dirac point ${\bf K}=\left(4\pi/3\sqrt{3},0\right)^T$, with strain-dependent gap $\Delta[\xi,\chi]$ shown in panel (d).}
\label{fig: phasediagram-topmag}
\end{figure}

We now show the existence of topological transitions for different classes of uniform strains. The topological nature of magnon eigenstates is captured by the Chern number $C^{\pm}=\int_{\text{BZ}}d{\bs k}\, \Omega^{\pm}_{\bs k}/2\pi$, where $\pm$ labels each magnonic band and the integration is over the first Brillouin zone (BZ). The Berry curvature is $\Omega^{\pm}_{\bs k}=\mp\hat{\bs h}\cdot({\partial_{k_x}\hat{\bs h}}\times{\partial_{k_y} \hat{\bs h}})/2$, with the unit vector $\hat{\bs h}={\bs h}/|{\bs h}|$ and the field ${\bs h}$ given by Eq. (\ref{eq: h-field}). Since we focus on uniform strains, we consider ${\Delta}_{\bs k\bs k'}={\Delta}_{\bs k}\delta_{\bs k\bs k'}$ and introduce the dimensionless parameters $\xi={\Delta^{(1)}}/{\Delta^{(2)}}$ and $\chi={\Delta^{(3)}}/{\Delta^{(2)}}$ to study the induced topological phases. The limit $\xi=\chi=1$ corresponds to gapped states at Dirac points, with energy gap $\Delta_g=3\sqrt{3}D_{\text{eff}}$ and Chern number $C^{\pm}=\pm 1$ for each magnon band. In the general case of uniform strains, i.e., $\xi\neq\chi\neq 1$, the topological character is preserved. We find for different lattice deformations described by the parameters $\xi$ and $\chi$, topological magnonic states experience a phase transition set by a sign change of the Chern number. Details of various strained configurations, with their respective Chern number, are found at SM. In Fig. \ref{fig: phasediagram-topmag}(a), the phase diagram for the topological phases is shown and featured by the respective Chern number $C^{+}(\xi,\chi)=-C^{-}(\xi,\chi)$. The transition that separates distinct topological phases is set when the Chern number nullifies, which corresponds to the closing of the strain-dependent magnon gap $\Delta_g[\xi,\chi]=\sqrt{3}D_{\text{eff}}(1+\xi+\chi)/{2}$ shown in panel (d). The field ${\bs h}$ and the magnon spectrum for each mode, $\epsilon_{\pm}({\bs k})=\Omega\pm\sqrt{{\bs h}_{\bs k}\cdot{\bs h}_{\bs k}}$, are plotted in panels (b) and (c), respectively. In the continuum limit and near Dirac points, the Hamiltonian \ref{eq: magnonHamiltonian} is ${H}^l_m=(2\pi)^{-2}\int d{\bs k}\Psi^{\dagger}_{\bs k}{\cal H}({\bs k})\Psi_{\bs k}$, where ${\cal H}({\bs k})=\Omega\tau_0+v({\bs k}\cdot{\bs \tau})+\left(\Delta_g+{\bs\Delta}\cdot {\bs k}\right)\tau_z$, $v=3sJ/2$ and ${\bs\Delta}=D_{\text{eff}}(-\sqrt{3}(1+\xi-2\chi),3(\xi-1))/4$. As a result, at the high-frequency limit, uniform strains break time-reversal and inversion symmetry, which provides a linear momentum mass and induces topologically nontrivial magnon states.

{\it Confined magnon states}.-- We now derive the continuum magnon theory that captures the effect of generic nonuniform strains. We start by representing lattice deformations as ${\bs u}={\bs u}_{l}+{\bs u}_{nl}$ with the assumption $|{\bs u}_{nl}|/|{\bs u}_{l}|<1$, where ${\bs u}_{nl}$ features displacements that depend nonlinearly with position. We find for the strain-induced and time-reversal symmetry broken term to satisfy $\Delta_{{\bs k}{\bs k}'}=\Delta_{\bs k}\delta_{{\bs k}{\bs k}'}+{\Delta}^{nl}_{{\bs k}{\bs k}'}$, with $\Delta_{\bs k}$ solely determined by the uniform strains. The nonlinear contribution $\Delta^{nl}_{{\bs k}{\bs k}'}=\sum_{\eta\eta'}\Delta^{(\eta\eta'),nl}_{{\bs k}{\bs k}'}e^{i{\bs k}'\cdot{\bs \delta}^{nn}_{\eta}}/N$ is determined by the magnon coupling $\Delta^{(\eta\eta'),nl}_{{\bs k}{\bs k}'}={\cal D}^{\eta\eta',-}_{\bs k-\bs k'}{\bs u}^{nl,+}_{\bs k-\bs k'}+{\cal D}^{\eta\eta',+}_{\bs k-\bs k'}{\bs u}^{nl,-}_{\bs k-\bs k'}$, which in turn depends on the Fourier components of the field of deformations ${\bs u}^{\pm}_{nl}$. In the last result, the operator in momentum space is given by ${\cal D}^{\eta\eta',\pm}_{\bs k}=\pm i(J')^2\left[(\nabla_{\eta}{\bs u}^{\pm}_l) {\bs k}_{\eta'}-(\nabla_{\eta'}{\bs u}^{\pm}_l) {\bs k}_{\eta}\right]$, which is dependent on the gradients of the fields ${\bs u}^{\pm}_l$.

Pairs of magnon states with different momenta ${\bs k}$ and ${\bs k}'$ become coupled as a result of applying nonuniform periodically varying strains. The effects on the topological gap and spectrum of states need to be evaluated perturbatively. Near the Dirac points, the interaction between magnons depends on their relative momentum difference. In turn, the magnon Hamiltonian \ref{eq: magnonHamiltonian} is written as
\begin{align}\label{eq: Dirac-magnon-Hamiltonian-nl}
{H}_D={H}^l_m+\int \frac{d{\bs k}}{(2\pi)^2}\frac{d{\bs k}'}{(2\pi)^2}\Psi^{\dagger}_{\bs k}\Delta^{nl}({\bs k-\bs k'})\tau_z\Psi_{\bs k'},
\end{align}
where the second term is the lowest-order correction in nonlinear deformation fields. Generally, an external periodically varying strain induces a massive term near the Dirac points. The effective mass $\Delta^{nl}({\bs x})$ is local in real space, breaking the continuous translational symmetry and significantly affecting the propagation of magnon states.

The energy spectrum of magnon excitations is obtained through the corresponding Dirac equation $\left[{\cal H}(-i{\bs \nabla})+\Delta^{nl}({\bs x})\tau_z\right]\Psi({\bs x},t)=i\partial_t\Psi({\bs x},t)$, where $\Psi=(\psi_+,\psi_-)^T$ is the two-component wavefunction for the magnon modes. {For the uniform component of the strain, it is assumed that general deformation fields are detailed at SM and captured by the parameters $\xi$ and $\chi$ introduced above.}
\begin{figure}[ht]
		\includegraphics[width=85mm]{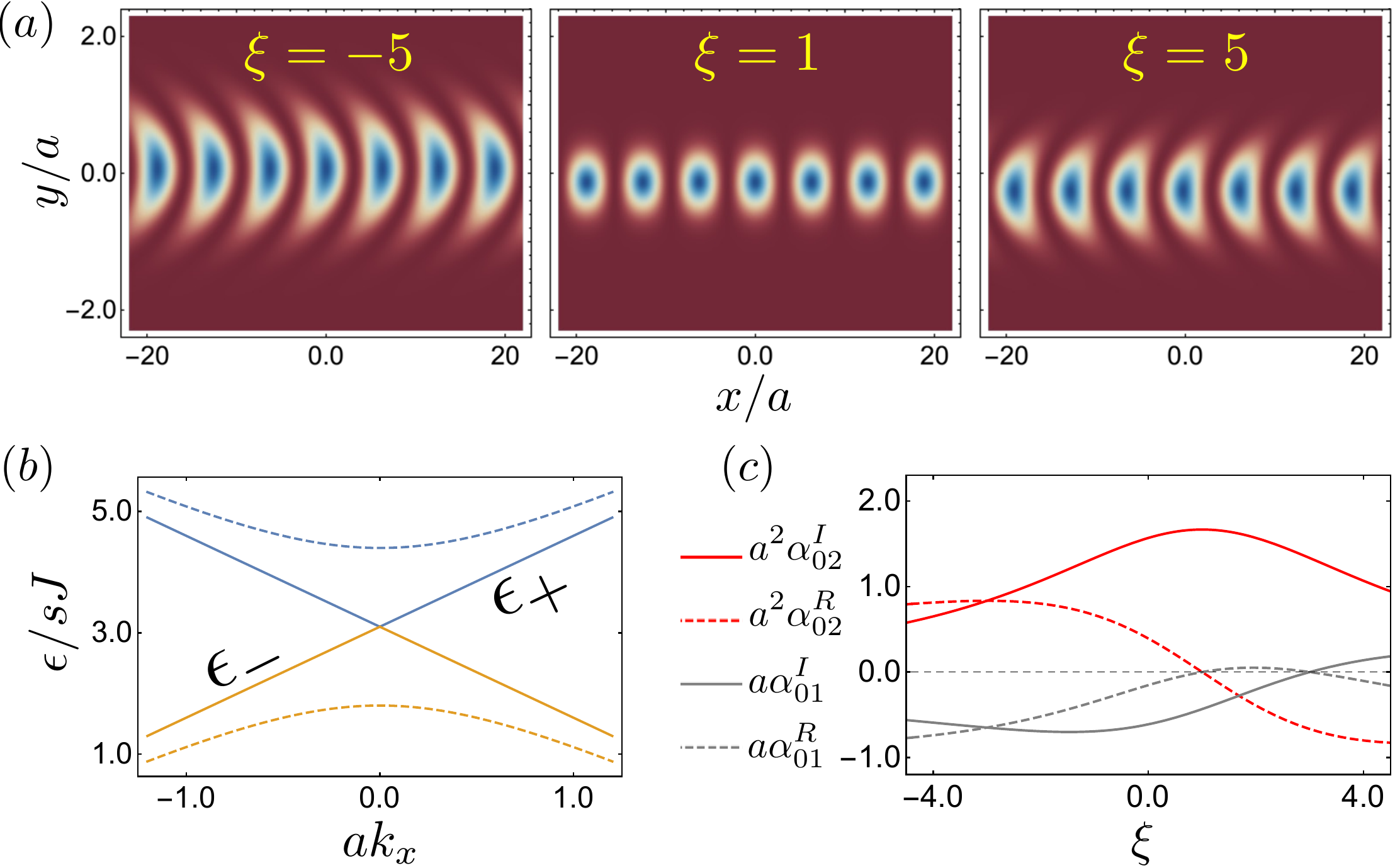}
		\caption{(a) Real-space distribution of confined magnon wave function for various strain parameters $\xi$. Nonuniform strain is applied along $y$-direction, which in turn determines direction of confinement. (b) Each mode propagates along $x$ direction with energy $\epsilon_{\pm}$ (solid lines), which are degenerate at Dirac points. In absence of nonlinear deformations, eigenstates are gapped (dashed line) for $\chi=\xi=1$. (c) Real and imaginary parts of coefficients $\alpha_{01}$ and $\alpha_{02}$ for $\chi=-2$ as a function of strain parameter $\xi$. Results in (a) and (b) are shown for parameters $\chi=1$, $\tilde{\beta}=10$ and $ak_x=0.5$, and $D_{\text{eff}}/sJ=0.5$, respectively.}
\label{fig: non-uniform strains}
\end{figure}
Furthermore, we assume a uniaxial nonuniform strain deduced from the field ${\bs u}^{\pm}_{nl}={\bs u}_{nl}=\beta y^2\hat{\bs y}$, with $\beta$ being the strength of the deformation. The effective mass results in $\Delta^{nl}({\bs x})=\tilde{\beta}y$, with $\tilde{\beta}={9\beta}(a_2+b_1)/2$ and the parameters $a_2$ and $b_1$ featuring the field ${\bs u}_l$. The magnonic Dirac equation is solved by the ansatz $\psi_{\pm}(\bs x,t)=A_{\pm}e^{ip({\bs x})}e^{-i\epsilon t/\hbar}$, where the polynomial $p({\bs x})=\sum_{nm}\alpha_{nm}x^ny^m$, quadratic on position coordinates, determines
the spatial dependence of the wavefunction. The equation for the eigenenergies, $\epsilon$, results in
\begin{align}\label{eq: eigenenergies-nonlinear}
\left[{\bs \Delta}\cdot{\bs \nabla}p+(\Delta+\Delta^{nl}({\bs x}))\right]^2+v^2({\bs\nabla}p)^2=\left(\Omega-\epsilon\right)^2,
\end{align}
which in turn produces a set of secular equations for $\alpha_{nm}$. The solution admits complex valued coefficients $\alpha_{01}$ and $\alpha_{02}$, where $\alpha_{20}=\alpha_{11}=0$ and $\alpha_{10}$ are also real valued; details can be found in the Supplemental Material. Therefore, the amplitude of the wavefunction is position dependent and given by ${\cal A}_{\pm}({\bs x})=A_{\pm}e^{-\text{Im}\left[p({\bs x})\right]}${ whose shape in space is determined by the parameters $\xi$ and $\chi$ from the nonuniform strain.} The spatial distribution of the magnon wavefunction diffuses along the direction of the applied strain, as shown in Fig. \ref{fig: non-uniform strains}(a), and its propagation becomes confined along the $x$ direction. The range of confinement is described by the coefficients $\alpha^I_{01}$ and $\alpha^I_{02}$, shown in Fig. \ref{fig: non-uniform strains}(c) as a function of the strain parameter $\xi$, which determines an exponential decay of the wavefunction along the $y$ direction. The energy of propagation is gapless and exhibits a linear dispersion around the Dirac points, $\epsilon_{\pm}=\Omega\pm v|k_x|/sJ$, as shown in panel (b). Note that in the absence of nonuniform strain, the amplitude of the wavefunction is homogeneous, and the energy around the Dirac point is gapped, as shown for $\chi=\xi=1$ by the dashed lines. Other types of deformations such as biaxial or triaxial strains might lead to interesting effects on the magnon spectrum; however, their analysis is left for future studies.

{\it Discussion and conclusions}.-- The assumption of time-dependent strains in the high-frequency range is at the core of the present theory. Thus, noticeable effects emerging from our model are expected in materials with a short bandwidth in their dispersion relation or, equivalently, a low Curie temperature.
Recent advances in manufacturing two-dimensional honeycomb magnetic materials with easy-axis anisotropy and low Curie temperature \cite{torelli2020high} would allow to demonstrate our predictions experimentally. Efforts to excite high-frequency phonons make our proposal feasible for future realizations through, for instance, considering a 2D magnetic material grown on a piezoelectric substrate that enables both spatial and temporal variations of deformation fields \cite{ZhaoPRL2022}.
Note that although the main results were presented for ferromagnets, our model admits generalizations that extend to other forms of magnetic order such as collinear antiferromagnets and magnetic textures.

In summary, we demonstrated the emergence of spin chirality in collinear honeycomb ferromagnets under high-frequency time-dependent strains. Given the limit of smoothly varying deformations, magnon fluctuations acquired an effective massive term that can be modified by the properties of strain, such as frequency and amplitude. Homogeneous strains induce a topological gap at the Dirac points and thus enable the control of topological magnonic phases. Interestingly, nonuniform uniaxial strains close the existing gap and confine the propagation of magnonic states due to the breaking of translational symmetry. The underlying physics of this phenomenon is different from systems under static strains, where elastic gauge fields emerge in response to smooth deformations. The ability to generate topological magnon phases and confine magnonic signals with time-dependent strains is of great interest in topological magnonics.


\begin{acknowledgments}
This work was supported by the Research Council of Norway through its Centres of Excellence funding scheme, Project No. 262633, ``QuSpin,'' and Fondecyt Iniciacion No. 11220046.
\end{acknowledgments}

\bibliography{manuscript}

\clearpage
\onecolumngrid
\section{Supplemental Material}
In this Supplemental Material, we explicitly show the calculation of the effective Hamiltonian by employing the Floquet theory for periodically driven quantum systems as well as relevant insights on linear and non-linear deformation fields.
\subsection{Effective magnon Hamiltonian}
We start with the time-dependent component of the Hamiltonian $V(t)=\sum_{\langle ij\rangle}{\cal J}_{i,j}{\bf S}_i\cdot{\bf S}_{j}$. By using that $S^{\pm}=S^x\pm i S^y$, $V(t)$ can be written as
\begin{align}
V(t)=\sum_{\langle ij\rangle}\left[{\cal J}^s_{ij}S^-_i{S}^+_j+{\cal J}_{ij}{S}^z_i{S}^z_j\right],
\end{align}
where we have labelled ${\cal J}^s_{ij}$ to the symmetric part of tensor ${\cal J}_{ij}$, whose notation (superindex $s$) will be dropped out hereafter. Let us consider an elliptically polarized time-dependent strain field as ${\bs u}_i(t)={\bs u}^a_{i}\sin \omega t+{\bs u}^b_i \cos{\omega t}$. Therefore, the matrix elements ${\cal J}_{ij}$ can be expanded as ${\cal J}_{ij}={\cal J}^{a}_{ij} \sin \omega t + {\cal J}^{b}_{ij}\cos{\omega t}$, where $\mathcal{J}_{ij=i+\delta_{\eta}}^{a,b} = J'{\bs \delta}_{\eta}\cdot\left({\bs \delta}\cdot\nabla\right){\bs u}_i^{a,b}$. Next, the Fourier component is $H_n=1/{\cal T}\int^{\cal T}_0 dt V(t)e^{in\omega t}$, where $\omega$ stands for the frequency of the drive and ${\cal T}$ the period of oscillations. Thus, by performing the proper integration we get
\begin{align}
H_n=\sum_{\langle ij\rangle}({\cal J}^{+}_{ij}\delta_{n,1}+{\cal J}^{-}_{ij}\delta_{n,-1}){S}^-_i{S}^+_j + \sum_{\langle ij\rangle}({\cal J}^{+}_{ij}\delta_{n,1}+{\cal J}^{-}_{ij}\delta_{n,-1}){S}^z_i{S}^z_j,
\end{align}
where we have introduced the definition $\mathcal{J}_{ij}^{\pm}=\left(\mathcal{J}_{ij}^b\pm i\mathcal{J}_{ij}^a\right)/2$. Since we are focused on the high-frequency limit, we employ the Brillouin-Wigner theory for periodically driven system to evaluate the effective Hamiltonian as an expansion in powers of $1/\omega$. In specific, the effective Hamiltonian is $H_{\text{eff}}=H+\sum_{n>0}\frac{\left[H_n,H_{-n}\right]}{n\omega}+\dots$ and, for our particular case, it reads 
\begin{align}
    H_{\text{eff}} = H + \frac{\left[H_1,H_{-1}\right]}{\omega}\delta_{n,1}.
\end{align}
By calculating explicitly the commutator, the effective Hamiltonian reads
\begin{align}
H_{\text{eff}} = H +\frac{2}{\omega}\sum_{\langle ij\rangle}\sum_{\langle jj'\rangle}\Delta_{ijj'}\left[S_j^z ( S_i^x S_{j'}^x + S_i^y S_{j'}^y ) + iS_j^z\bs{\hat{z}} \cdot ( \bs{S}_i \times \bs{S}_{j'} )+S_i^z ( S_{j'}^x S_{j}^x + S_{j'}^y S_{j}^y ) + S_{j'}^z ( S_{j}^x S_{i}^x + S_{j}^y S_{i}^y )\right],
\end{align}
with $\Delta_{ijj'}={\cal J}^{+}_{ij}{\cal J}^{-}_{jj'}-{\cal J}^{+}_{jj'}{\cal J}^{-}_{ij}$. By employing symmetry arguments on $\Delta_{ijj'}$, we finally arrive to
\begin{align}
    H_{\text{eff}}=H+\frac{2i}{\omega}\sum_{\langle ij\rangle}\sum_{\langle jj'\rangle}\Delta_{ijj'}\,
{\bs S}_j\cdot \left( \bs{S}_i \times \bs{S}_{j'}\right),
\end{align}
that corresponds to Eq. (\ref{eq:effecHam}) of the main text. 
\subsection{Linear deformations}
In order to study the magnon excitation in the strained system, we first consider linear deformations. Recall that the mechanical degrees of freedom are encoded in the field $\Delta_{{\bs k}{\bs k}'}$ that is proportional to the $z-$component of ${\bs h}_{{\bs k}{\bs k}'}$ at Eq. (\ref{eq: h-field}) in the main text. It will be useful to define the nearest neighbor vectors:
\begin{align}
    \bs \delta_1=&a\left(\sqrt{3}/2,-1/2,0\right)\\
    \bs \delta_2=&a\left(0,1,0\right)\\
    \bs \delta_3=&-a\left(\sqrt{3}/2,1/2,0\right),
\end{align}
and also the next-nearest neighbors vectors 
\begin{align}
    \bs \delta_1^{nn}= {\bs \delta}_1 - {\bs \delta}_3= - \bs \delta_4^{nn}\\
    \bs \delta_2^{nn}= {\bs \delta}_2 - {\bs \delta}_1= - \bs \delta_5^{nn}\\
    \bs \delta_3^{nn}= {\bs \delta}_3 - {\bs \delta}_2= - \bs \delta_6^{nn}
\end{align}

The non-local field $\Delta_{{\bs k}{\bs k}'}$ is explicitly given by
\begin{align}
\label{deltakk'}
  \Delta_{{\bs k}{\bs k}'}
=\frac{2 i}{N}\sum_{\eta}\Delta_{{\bs k}{\bs k}'}^{\eta}\sin\left(\bs{k}'\cdot\bs{\delta}^{nn}_{\eta}\right),
\end{align}
with
$\Delta_{{\bs k}{\bs k}'}^{\eta}=\sum_{i}\Delta_{i,i-\bs{\delta}_{\eta},i-\bs{\delta}^{nn}_{\eta}} e^{-i({\bs k}-{\bs k}')\cdot{\bs r}_i}$. In the linear regime, $\Delta_{{\bs k}{\bs k}'}^{\eta}$ is local in the momentum space since $\Delta_{i,i-\bs{\delta}_{\eta},i-\bs{\delta}^{nn}_{\eta}}$ becomes independent of position. Therefore, $\Delta_{{\bs k}{\bs k}'}^{\eta}=\Delta_{i,i-\bs{\delta}_{\eta},i-\bs{\delta}^{nn}_{\eta}}\delta_{{\bs k}{\bs k'}}$. The next step is to consider some explicit deformation field to evaluate the field $\Delta_{i,i-\bs{\delta}_{\eta},i-\bs{\delta}^{nn}_{\eta}}$. Let us write the strain fields in the most general form $
    \bs{u}_i^a = \alpha_1^a(x,y)\bs{\hat{x}}+\alpha_2^a(x,y)\bs{\hat{y}}$ and $\bs{u}_i^b = \alpha_1^b(x,y)\bs{\hat{x}}+\alpha_2^b(x,y)\bs{\hat{y}}$, with the coefficients given by
\begin{align}
    \alpha_1^a(x,y) &= a_1x + a_2y,\\
    \alpha_2^a(x,y) &= b_1x + b_2y,\\
    \alpha_1^b(x,y) &= c_1x + c_2y,\\
    \alpha_2^b(x,y) &= d_1x + d_2y.
\end{align}
This, allows us to write $\Delta_{{\bs k}{\bs k}'}=\Delta_{\bs k}\delta_{{\bs k}{\bs k}'}$, with 
\begin{align}
\Delta_{\bs k}=-a^4\sum_{\eta}\Delta^{(\eta)}\sin\left({\bs k}\cdot{\bs \delta}_{\eta}^{nn}\right),
\end{align}
and
\begin{align}
    \Delta^{(1)}&=3a_1d_2-\sqrt{3}d_2 (a_2+b_1)-3 b_2 c_1+\sqrt{3} b_2 (c_2+d_1)\\
    \Delta^{(2)}&=-3 a_1 d_2-\sqrt{3} d_2 (a_2+b_1)+3 b_2 c_1+\sqrt{3} b_2 (c_2+d_1)\\
    \Delta^{(3)}&=\sqrt{3} \left[(a_2+b_1) (3 c_1+d_2)-(3 a_1+b_2) (c_2+d_1)\right].
\end{align}
In particular, by imposing that $\Delta^{(1)}=\Delta^{(2)}=\Delta^{(3)}=\Delta$, a possible solution reads $\Delta = \frac{3}{2} \sqrt{3}c_1 (a_2+b_1)$ which is proportional to the effective Dzyaloshinskii-Moriya interaction $D_{\text{eff}}=-{4s^2a^4(J')^2\Delta}/{\omega}$, as stated in the main text.
\subsubsection{Particular solutions}
Here we show some particular examples of deformation fields which produce non-trivial topological magnon states. Since our parametrization allows several solutions for deformation fields, we show explicitly some characteristic cases with the Chern numbers $C^{-}$, $C^{+}$ and also the noticeable result that mimics the DM interaction through $D_{\text{eff}}$ (also $C^{+}$).
\begin{figure}[ht]
		\includegraphics[width=170mm]{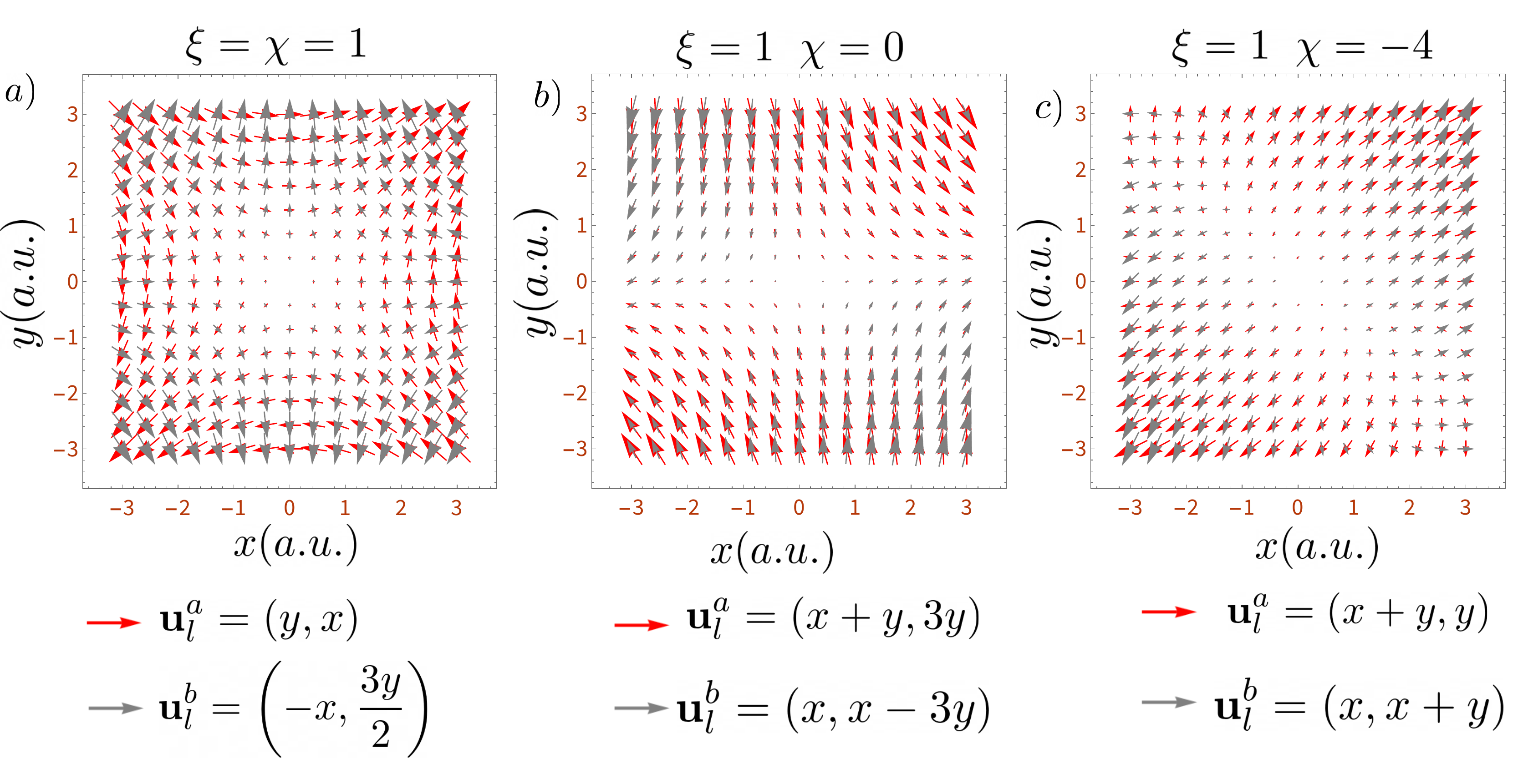}
		\caption{Vector representation of deformation fields  $\mathbf{u}_l^a$ and  $\mathbf{u}_l^b$ for different values of the $\chi$ and $\xi$. Panel a) corresponds to $\chi=\xi=1$ with Chern number $C^{+}$, b) $\xi=1, \chi=0$ with Chern number $C^{+}$, and c) $\xi=1, \chi=-4$ with Chern numner $C^{-}$. In all cases the red and gray arrows represent the deformation fields $\mathbf{u}_l^a$ and $\mathbf{u}_l^b$, respectively.}
\label{fig: SM}
\end{figure}
Fig. \ref{fig: SM} shows in panels a), b), and c) the deformation fields for the cases $\chi=\xi=1$; $\chi=0$, $\xi = 1$; and $\chi=1$, $\xi = -4$. The first one corresponds to the highlighted effective DM interaction with Chern number $C^{-}$, while the two latter correspond to $C^{-}$ and $C^{+}$. At the bottom of each vector representation is shown the respective expression for the vector fields $\mathbf{u}_l^a$ and  $\mathbf{u}_l^b$.

\subsection{Non-linear deformations}
The time-dependent part of Hamiltonian (\ref{eq: magnonHamiltonian}) can be written as
\begin{align}
H'=(J')^2\frac{4s^2}{\omega}\sum_{{\bs k},{\bs k}'}{\Delta}_{\bs k\bs k'}\psi^{\dagger}_{\bs k}\tau_z\psi_{\bs k'}.
\end{align}
where
\begin{align}
\Delta_{{\bs k}{\bs k}'}=\frac{2 i}{N}\left[\Delta_{{\bs k}{\bs k}'}^{12}\sin\left(\bs{k}'\cdot(\bs{\delta}_1 - \bs{\delta}_2)\right)+\Delta_{{\bs k}{\bs k}'}^{31}\sin\left(\bs{k}'\cdot(\bs{\delta}_3 - \bs{\delta}_1)\right)+\Delta_{{\bs k}{\bs k}'}^{23}\sin\left(\bs{k}'\cdot(\bs{\delta}_2 - \bs{\delta}_3)\right)\right].
\end{align}

Let us consider the inhomogeneous deformation field ${\bs u}({\bs r})$, being non-linear on their dependence with the position. As first approximation, we assume the non-linear contribution a slight deviation from the linear contribution,
\begin{align}
{\bs u}({\bs r})={\bs u}_{l}({\bs r})+{\bs u}_{nl}({\bs r})
\end{align}
where $|{\bs u}_{l}|/|{\bs u}_{nl}|<1$. In this approximation, we find $\Delta_{{\bs k}{\bs k}'}=\Delta_{\bs k}\delta_{{\bs k}{\bs k}'}+{\Delta}^{nl}_{{\bs k}{\bs k}'}$, where ${\Delta}^{nl}_{{\bs k}{\bs k}'}$ is to be determined. 

Back to the definition of $\Delta_{i,i-\delta_{\eta},i-\delta_{\eta}^{nn}}$ (see definition of $\Delta_{\bs k k'}^{\eta}$ in Eq. ({\ref{deltakk'}})), we invoke the mentioned approximation on the deformation field to find $\Delta_{i,i-\delta_{\eta},i-\delta_{\eta}^{nn}}\approx\Delta^{l}_{i,i-\delta_{\eta},i-\delta_{\eta}^{nn}}+\Delta^{nl}_{i,i-\delta_{\eta},i-\delta_{\eta}^{nn}}$, where quadratic contributions on ${\bs u}_{nl}$ has been dropped. The expression for  $\Delta^{nl}$ reads
\begin{align}
\Delta&^{nl}_{i,i-\delta_{\eta},i-\delta_{\eta}^{nn}}=\\
&+(J')^2\left[{\bs \delta}_{\eta}\cdot({\bs \delta}_{\eta}\cdot\nabla){\bs u}^{nl,+}_{i} {\bs \delta}_{\eta'}\cdot({\bs \delta}_{\eta'}\cdot\nabla){\bs u}^{l,-}_{i}-{\bs \delta}_{\eta}\cdot({\bs \delta}_{\eta}\cdot\nabla){\bs u}^{l,-}_{i} {\bs \delta}_{\eta'}\cdot({\bs \delta}_{\eta'}\cdot\nabla){\bs u}^{nl,+}_{i}\right]\\
&+(J')^2\left[{\bs \delta}_{\eta}\cdot({\bs \delta}_{\eta}\cdot\nabla){\bs u}^{l,+}_{i} {\bs \delta}_{\eta'}\cdot({\bs \delta}_{\eta'}\cdot\nabla){\bs u}^{nl,-}_{i}-{\bs \delta}_{\eta}\cdot({\bs \delta}_{\eta}\cdot\nabla){\bs u}^{nl,-}_{i} {\bs \delta}_{\eta'}\cdot({\bs \delta}_{\eta'}\cdot\nabla){\bs u}^{l,+}_{i}\right]\\
&=(J')^2\left[ {\bs \delta}_{\eta'}\cdot({\bs \delta}_{\eta'}\cdot\nabla){\bs u}^{l,-}_{i}\left({\bs \delta}_{\eta}\cdot({\bs \delta}_{\eta}\cdot\nabla)\right)-{\bs \delta}_{\eta}\cdot({\bs \delta}_{\eta}\cdot\nabla){\bs u}^{l,-}_{i} \left({\bs \delta}_{\eta'}\cdot({\bs \delta}_{\eta'}\cdot\nabla)\right)\right]{\bs u}^{nl,+}_{i}\\
&+(J')^2\left[{\bs \delta}_{\eta}\cdot({\bs \delta}_{\eta}\cdot\nabla){\bs u}^{l,+}_{i} \left({\bs \delta}_{\eta'}\cdot({\bs \delta}_{\eta'}\cdot\nabla)\right)- {\bs \delta}_{\eta'}\cdot({\bs \delta}_{\eta'}\cdot\nabla){\bs u}^{l,+}_{i}\left({\bs \delta}_{\eta}\cdot({\bs \delta}_{\eta}\cdot\nabla)\right)\right]{\bs u}^{nl,-}_{i},
\end{align}
which is compactly written as
\begin{align}\label{eq:deltanonlinear-re}
\Delta^{nl}_{i,i-\delta_{\eta},i-\delta_{\eta}^{nn}}={\cal D}^{-}_{\eta\eta'}{\bs u}^{nl,+}_{i}+{\cal D}^{+}_{\eta\eta'}{\bs u}^{nl,-}_{i}.
\end{align}

Note that ${\cal D}^{\pm}_{\eta\eta'}$ are linear operators acting on the fields ${\bs u}^{nl,\mp}_{i}$, respectively. These operators become trivial in the absence of linear components of the strain. In Fourier space we set ${\bs u}^{nl,\pm}_{i}=\sum_{\bs q}{\bs u}^{nl,\pm}_{\bs q}e^{i{\bs q}\cdot{\bs r}_i}$ and thus, the action of the operators becomes ${\cal D}^{\mp}_{\eta\eta'}{\bs u}^{nl,\pm}_{i}=i\sum_{\bs q}e^{i{\bs q}\cdot{\bs r}_i}{\cal D}^{\eta\eta',\mp}_{\bs q}{\bs u}^{nl,\pm}_{\bs q}$, where 
\begin{align}
{\cal D}^{\eta\eta',-}_{\bs q}=(J')^2\left[ {\bs \delta}_{\eta'}\cdot({\bs \delta}_{\eta'}\cdot\nabla){\bs u}^{l,-}_{i}\left({\bs \delta}_{\eta}\cdot({\bs \delta}_{\eta}\cdot {\bs q})\right)-{\bs \delta}_{\eta}\cdot({\bs \delta}_{\eta}\cdot\nabla){\bs u}^{l,-}_{i} \left({\bs \delta}_{\eta'}\cdot({\bs \delta}_{\eta'}\cdot {\bs q})\right)\right].
\end{align}

Therefore, an expression for $\Delta^{\eta\eta',nl}_{{\bs k}{\bs k}'}$, using Eq. (\ref{eq:deltanonlinear-re}), is found to be
\begin{align}
\Delta^{\eta\eta',nl}_{{\bs k}{\bs k}'}=\sum_{i\in{\cal B}}\Delta^{nl}_{i,i-\delta_{\eta},i-\delta_{\eta}^{nn}} e^{-i({\bs k}-{\bs k}')\cdot{\bs r}_i}=i{\cal D}^{\eta\eta',-}_{\bs k-\bs k'}{\bs u}^{nl,+}_{\bs k-\bs k'}+i{\cal D}^{\eta\eta',+}_{\bs k-\bs k'}{\bs u}^{nl,-}_{\bs k-\bs k'}.
\end{align}
Finally, we ended up with a short expression for the non-linear contribution to the field $\Delta_{{\bs k}{\bs k}'}$, defined by $\Delta^{nl}_{{\bs k}{\bs k}'}=\frac{1}{N}\sum_{\eta\eta'}\Delta^{\eta\eta',nl}_{{\bs k}{\bs k}'}e^{i{\bs k}'\cdot\left({\bs \delta}_{\eta}-{\bs \delta}_{\eta'}\right)}$, ($N$ the number of lattice sites) and that satisfies
\begin{align}
\Delta^{nl}_{{\bs k}{\bs k}'}=\frac{i}{N}\sum_{\eta\eta'}\left({\cal D}^{\eta\eta',-}_{\bs k-\bs k'}{\bs u}^{nl,+}_{\bs k-\bs k'}+{\cal D}^{\eta\eta',+}_{\bs k-\bs k'}{\bs u}^{nl,-}_{\bs k-\bs k'}\right)e^{i{\bs k}'\cdot\left({\bs \delta}_{\eta}-{\bs \delta}_{\eta'}\right)},
\end{align}
which might also be written as:
\begin{align}\label{eq:Deltakk'}
\Delta^{nl}_{{\bs k}{\bs k}'}=\frac{2 i}{N}\left[\Delta_{{\bs k}{\bs k}'}^{12,nl}\sin\left(\bs{k}'\cdot(\bs{\delta}_1 - \bs{\delta}_2)\right)+\Delta_{{\bs k}{\bs k}'}^{31,nl}\sin\left(\bs{k}'\cdot(\bs{\delta}_3 - \bs{\delta}_1)\right)+\Delta_{{\bs k}{\bs k}'}^{23,nl}\sin\left(\bs{k}'\cdot(\bs{\delta}_2 - \bs{\delta}_3)\right)\right]
\end{align}
The total Hamiltonian including non-linear deformations now reads, 
\begin{align}
{H}=\sum_{\bs k}\Psi^{\dagger}_{\bs k}\left(\Omega\mathbb{I}+{\bs h}_{\bs k}\cdot{\bs{\tau}}\right)\Psi_{\bs k}+(J')^2\frac{4s^2}{\omega}\sum_{\bs k\bs k'}\Psi^{\dagger}_{\bs k}\Delta^{nl}_{\bs k\bs k'}{{\tau}_z}\Psi_{\bs k'}.
\end{align}
In a small window, near the Dirac point ${\bf K}_{+}=\left(4\pi/3\sqrt{3},0\right)^T$, we find for the magnon Hamiltonian
\begin{align}\label{eq:DiracHam1}
{H}_D=\sum_{\bs k}\Psi^{\dagger}_{\bs k}\left[\Omega\tau_0+v\left(k_x\tau_x+k_y\tau_y\right)+\left(\Delta_g+{\bs\Delta}\cdot {\bs k}\right)\tau_z\right]\Psi_{\bs k}+\sum_{\bs k\bs k'}\Psi^{\dagger}_{\bs k}\Delta^{nl}_{\bs k\bs k'}{{\tau}_z}\Psi_{\bs k'},
\end{align}
where $\Delta_g(\xi,\chi)=\sqrt{3}D_{\text{eff}}(1+\xi+\chi)/{2}$, $v=3sJ/2$, and the vector ${\bs\Delta}=D_{\text{eff}}(-\sqrt{3}(1+\xi-2\chi),3(\xi-1))/4$. To find $\Delta^{nl}_{\bs k\bs k'}$ near the Dirac points, we first note from Eq. (\ref{eq:Deltakk'})
\begin{align}
\sin\left(\bs{k}\cdot(\bs{\delta}_1 - \bs{\delta}_2)\right)&\approx\frac{\sqrt{3}}{2}+{\cal O}(\bs{k}),\\
\sin\left(\bs{k}\cdot(\bs{\delta}_3 - \bs{\delta}_1)\right)&\approx\frac{\sqrt{3}}{2}+{\cal O}(\bs{k}),\\
\sin\left(\bs{k}\cdot(\bs{\delta}_2 - \bs{\delta}_3)\right)&\approx\frac{\sqrt{3}}{2}+{\cal O}(\bs{k}).
\end{align}

In turn, the field $\Delta^{nl}_{\bs k\bs k'}$ becomes dependent on ${\bs k - \bs k'}$, since
\begin{align}\label{eq:deltanlmomentum}
\Delta^{nl}_{{\bs k}{\bs k}'}\approx\frac{\sqrt{3}}{2}\frac{2 i}{N}\left[\Delta_{{\bs k}{\bs k}'}^{12,nl}+\Delta_{{\bs k}{\bs k}'}^{31,nl}+\Delta_{{\bs k}{\bs k}'}^{23,nl}\right],
\end{align}
with $\Delta^{\eta\eta',nl}_{{\bs k}{\bs k}'}=i{\cal D}^{\eta\eta',-}_{\bs k-\bs k'}{\bs u}^{nl,+}_{\bs k-\bs k'}+i{\cal D}^{\eta\eta',+}_{\bs k-\bs k'}{\bs u}^{nl,-}_{\bs k-\bs k'}$. Thus, we write the last term in Eq. (\ref{eq:DiracHam1}) as
\begin{align}
\sum_{\bs k\bs k'}\Psi^{\dagger}_{\bs k}\Delta^{nl}_{\bs k\bs k'}{{\tau}_z}\Psi_{\bs k'}&\approx\int \frac{d{\bs k}}{(2\pi)^2}\frac{d{\bs k}'}{(2\pi)^2}\Psi^{\dagger}({\bs k})\Delta^{nl}({\bs k-\bs k'}){{\tau}_z}\Psi(\bs k')\\
&=\int d{\bs r}\int \frac{d{\bs k}}{(2\pi)^2}\frac{d{\bs k}'}{(2\pi)^2}\Psi^{\dagger}({\bs k})\Delta^{nl}({\bs r})e^{-i({\bs k}-{\bs k}')\cdot{\bs r}}{{\tau}_z}\Psi(\bs k')\\
&=\int d{\bs r}\Psi^{\dagger}({\bs r})\Delta^{nl}({\bs r}){{\tau}_z}\Psi(\bs r),
\end{align}
which corresponds to Eq. (\ref{eq: Dirac-magnon-Hamiltonian-nl}) of the main text.

To explore the effects from the field $\Delta^{nl}({\bs r})$, let us assume an uniaxial strain induced by the deformation field ${\bs u}^{nl}=u_x\hat{\bs r}+u_y\hat{\bs y}$, with
\begin{align}
u_x&=0,\\
u_y&=\beta y^2,
\end{align}
and where the parameter $\beta$ denotes the strength of the strain. For the sake of simplicity, we consider ${\bs u}^{nl}_a=0$, which in turn imply ${\bs u}^{nl,\pm}={\bs u}^{nl}_b={\bs u}^{nl}$. Under this assumption we can simplify the operator at Eq. (\ref{eq:deltanonlinear-re}), which in the continuum limit reads,
\begin{align}\label{eq: DeltaOperator}
\Delta^{\eta\eta',nl}({\bs r})=\left({\cal D}^{-}_{\eta\eta'}+{\cal D}^{+}_{\eta\eta'}\right){\bs u}^{nl}=2i\left[{\cal D}_{\eta}{\bs u}^a_{l}{\cal D}_{\eta'}-{\cal D}_{\eta'}{\bs u}^a_{l}{\cal D}_{\eta}\right]{\bs u}^{nl},
\end{align}
with ${\bs u}^{a,b}_{l}$ characterizing the linear component of strain and ${\cal D}_{\eta}={\bs \delta}_{\eta}\cdot({\bs \delta}_{\eta}\cdot\nabla)$. The field $\Delta^{nl}({\bs r})$ is determined through its Fourier transform, Eq. (\ref{eq:deltanlmomentum}), which is defined by
\begin{align}
\Delta^{nl}({\bs q})=\frac{i\sqrt{3}}{N}\left[\Delta_{{\bs q}}^{12,nl}+\Delta_{{\bs q}}^{31,nl}+\Delta_{{\bs q}}^{23,nl}\right],
\end{align}
and thus
\begin{align}
\Delta^{nl}({\bs r})&\nonumber=\int \frac{d{\bs q}}{(2\pi)^2}\Delta^{nl}({\bs q})e^{i{\bs q}\cdot{\bs r}}\\
&=\frac{i\sqrt{3}}{N}\left(\Delta^{12,nl}({\bs r})+\Delta^{31,nl}({\bs r})+\Delta^{23,nl}({\bs r})\right),
\end{align}
with the components $\Delta^{\eta\eta',nl}({\bs r})$ obtained from Eq. (\ref{eq: DeltaOperator}). In summary, for the specific uniaxial non-uniform strain, the non-linear contribution adopts a simple form denoted by $\Delta^{nl}({\bs r})= \tilde{\beta}y$, with $\tilde{\beta} =\frac{9\beta}{2}(a_2+b_1)$

\subsection{Eigenvalues}

The equation of motion for the magnon field $\Psi(\bs r,t)=(\psi_{\alpha}(\bs r)e^{-i\epsilon t},\psi_{\beta}(\bs r)e^{-i\epsilon t})^T$, obtained in real space can be expressed in a matrix form as  
\begin{align}\label{eq:diraceq}
\left(\begin{array}{cc}
 \Omega+\Delta_g-i{\bs \Delta}\cdot\nabla +\Delta^{nl}({\bs r})    & -iv\left(\partial_x-i\partial_y\right)\\
 -iv\left(\partial_x+i\partial_y\right)   & \Omega-\Delta_g+i{\bs \Delta}\cdot\nabla-\Delta^{nl}({\bs r})  
\end{array}\right)\left(\begin{array}{c}
    \psi_{\alpha}  \\
    \psi_{\beta} 
\end{array}\right)=\epsilon\left(\begin{array}{c}
    \psi_{\alpha}   \\
    \psi_{\beta}
\end{array}\right).
\end{align}
Note that Eq. (\ref{eq:diraceq}) is a general result and can be evaluated both in the presence or absence of mechanical deformations.
\subsubsection{Solutions}
{\bf Case 1:} In order to solve Eq. (\ref{eq:diraceq}), we choose the anzats $\psi_{\alpha,\beta}({\bs r})=A_{\alpha,\beta}e^{i{\bs q}\cdot{\bs r}}$. In the absence of strains, Eq. (\ref{eq:diraceq}) reduces to 
\begin{align}
\left(\begin{array}{cc}
 \Omega-\epsilon    & v\left(q_x-iq_y\right)\\
 v\left(q_x+iq_y\right)   & \Omega-\epsilon
\end{array}\right)\left(\begin{array}{c}
    A_{\alpha}  \\
    A_{\beta} 
\end{array}\right)
=\left(\begin{array}{c}
    0   \\
    0
\end{array}\right),
\end{align}
with the eigenvalues given by $\epsilon=\Omega\pm v|{\bs q}|$. 

{\bf Case 2:} In presence of linear deformation fields, Eq. (\ref{eq:diraceq}) reduces to
\begin{align}
\left(\begin{array}{cc}
 \Omega-\epsilon+\Delta_g +{\bs \Delta}\cdot {\bs q}   & v\left(q_x-iq_y\right)\\
 v\left(q_x+iq_y\right)   & \Omega-\epsilon-\Delta_g-{\bs \Delta}\cdot {\bs q}
\end{array}\right)\left(\begin{array}{c}
    A_{\alpha}  \\
    A_{\beta} 
\end{array}\right)
=\left(\begin{array}{c}
    0   \\
    0
\end{array}\right),
\end{align}
with the eigenvalues given by
\begin{align}
\epsilon=\Omega\pm\sqrt{v^2|{\bs q}|^2+(\Delta_g+{\bs \Delta}\cdot {\bs q})^2}
\end{align}

{\bf Case 3:} In presence of a non-linear deformation field we generalizes the anzats by $\psi_{\alpha,\beta}({\bs r})=A_{\alpha,\beta}e^{ip({\bs r})}$, with the quadratic polynomial 
$p({\bs r})=\sum_{nm}\alpha_{nm}x^ny^m$. We assume that $\alpha_{10}$ is real valued, while $\alpha_{20}$, $\alpha_{02}$, $\alpha_{01}$, and $\alpha_{11}$ can be complex numbers. Thus, the set of equations become,
\begin{align}
\left(\Omega-{\epsilon}+\Delta_g+\Delta^{nl}({\bs r})\right)A_{\alpha}&=-{\bs \Delta}\cdot{\bs \nabla}  p({\bs r})A_{\alpha}-v(\partial_xp({\bs r})-i\partial_yp({\bs r}))A_{\beta}\\
\left(\Omega-{\epsilon}-\Delta_g-\Delta^{nl}({\bs r})\right)A_{\beta}&=-v(\partial_xp({\bs r})+i\partial_yp({\bs r}))A_{\alpha}+{\bs \Delta}\cdot{\bs \nabla} p({\bs r})A_{\beta}
\end{align}

If $A_{\alpha,\beta}\neq 0$, these equations can be combined to produce,
\begin{align}\label{eq: eigenenergies-nonlinearappen}
\left[{\bs \Delta}\cdot{\bs \nabla}p+(\Delta_g+\Delta^{nl}({\bs r}))\right]^2+v^2({\bs\nabla}p)^2=\left(\Omega-\epsilon\right)^2,
\end{align}
which corresponds to Eq. (\ref{eq: eigenenergies-nonlinear}) of the main text. Assuming the previous form for the polynomial $p({\bs r})$, the previous relation is expressed as $0=f_0+f_xx+f_yy+f_{yy}y^2+f_{xx}x^2+f_{xy}xy$. In turn, we find the following secular equations for the coefficients,
%
%
\begin{align}
f_0&=(\Omega-i\alpha_{02})^2-v^2(\alpha_{10}^2+\alpha_{01}^2)-(\Delta_g+\alpha_{10}\Delta_x+\alpha_{01}\Delta_y)^2=0\\
f_y&=-2v^2(2\alpha_{01}\alpha_{02}+\alpha_{10}\alpha_{11})-2(\Delta_g+\alpha_{10}\Delta_x+\alpha_{01}\Delta_y)(\Delta^{nl}+2\Delta_y\alpha_{02}+\Delta_x\alpha_{11})=0\\
f_x&=-2v^2(2\alpha_{10}\alpha_{20}+\alpha_{01}\alpha_{11})-2(\Delta_g+\alpha_{10}\Delta_x+\alpha_{01}\Delta_y)(2\Delta_x\alpha_{20}+\Delta_y\alpha_{11})=0\\
f_{xy}&=-2(\Delta^{nl}+2\Delta_y\alpha_{02}+\Delta_x\alpha_{11})(2\Delta_x\alpha_{20}+\Delta_y\alpha_{11})-4v^2(\alpha_{20}+\alpha_{02})\alpha_{11}=0\\
f_{yy}&=-(\Delta^{nl}+2\Delta_y\alpha_{02}+\Delta_x\alpha_{11})^2-v^2(4\alpha_{02}^2+\alpha_{11}^2)=0\\
f_{xx}&=-(2\Delta_x\alpha_{20}+\Delta_y\alpha_{11})^2-v^2(4\alpha_{20}^2+\alpha_{11}^2)=0,
\end{align}
where $\Delta_{x,y}$ denotes the $x-$ and $y-$ component of vector ${\bs \Delta}$, while $\Delta_g$ stands for the strain-dependent magnon gap. Note that all the coefficients depend on the parameters $\chi$ and $\xi$ through $\Delta_{x(y)}$ and $\Delta^{nl}$, which encode the external strain.
A simple and particular solution can be obtained when $\alpha_{20}=\alpha_{11}=0$,
\begin{align}
\psi_{\alpha}=A_{\alpha}e^{i(\alpha_{10} x+\alpha_{01}^Ry-\alpha_{02}^Ry^2)}e^{-\alpha_{01}^Iy+\alpha_{02}^Iy^2},
\end{align}
where the super index $R(I)$ stands for the real(imaginary) part of the respective coefficient. Thus, the amplitude of the magnon wavefunction reads ${\cal A}_{\pm}({\bs r})=A_{\pm}e^{-\text{Im}\left[p({\bs r})\right]}$, which is properly showed at panel (c) of Fig. \ref{fig: non-uniform strains} in the main text, and the two magnon bands in the presence of non-linear deformation fields reads $\epsilon_{\pm}=\Omega\pm v|k_x|$.
\end{document}